\documentclass[prb,twocolumn,amsmath,amssymb,superscriptaddress,longbibliography]{revtex4-2}
\usepackage[T1]{fontenc}

\usepackage{amsmath}
\usepackage{amsthm}
\usepackage{amssymb}
\usepackage{graphicx}
\usepackage{float}
\usepackage{hyperref}
\usepackage{braket}
\usepackage[utf8]{inputenc}
\usepackage{placeins}

\newcommand{\gsum}{\Sigma}

\usepackage[english]{babel}
\begin{document}

\title{Finite-time quenches across first-, second-, and infinite-order transitions in XXZ spin chains}

\author{Luiz Felipe Santos Dittberner}%
\affiliation{Instituto de F\'{i}sica and International Center for Physics, Universidade de Bras\'{i}lia, Bras\'{i}lia 70919-970, DF, Brazil}%
\author{Luigi Berber}%
\affiliation{Instituto de F\'{i}sica and International Center for Physics, Universidade de Bras\'{i}lia, Bras\'{i}lia 70919-970, DF, Brazil}%
\author{Maria C. O. Aguiar}%
\affiliation{Departamento de F\'isica, Universidade Federal de Minas Gerais, C.P.~702, 30123-970, Belo Horizonte, MG, Brazil}%
\author{Krissia Zawadzki}
\affiliation{Instituto de F\'{i}sica de S\~ao Carlos, Universidade de S\~ao Paulo,
CP 369, 13560-970 S\~ao Carlos, S\~ao Paulo, Brazil}
\author{Helena Bragan\c{c}a}%
\affiliation{Instituto de F\'{i}sica and International Center for Physics, Universidade de Bras\'{i}lia, Bras\'{i}lia 70919-970, DF, Brazil}%
  
\date{\today}

\begin{abstract}
We investigate how equilibrium quantum phase transitions manifest themselves in small systems driven out of equilibrium by quantum quenches. We consider linear ramps of the anisotropy in spin-1/2 and spin-1 XXZ chains, which host transitions of first, second, and infinite (Berezinskii–Kosterlitz–Thouless, BKT) order. We consider system sizes available in current quantum simulators and investigate how the order of the transition, the symmetries, and the ramp duration affect the decomposition of the time-evolved state into the instantaneous eigenbasis. The order of the transition determines the dynamics: across the BKT transition, moderately slow ramps already keep the system close to its ground state, corresponding to an adiabatic evolution, whereas for the second-order Haldane–Néel transition a ramp roughly three times slower is required to reach the same regime. Across the first-order transition, since the dynamics conserves the magnetization of the initial state, we observe the formation of ferromagnetic domains, separated by a single domain wall for slow ramps and by multiple walls for faster ones. Interestingly, shortly after the transition, the state evolves as a coherent beating between domain-wall states, a behavior captured by the recently proposed quantum-state texture.
\end{abstract}

\maketitle

\section{Introduction}

The non-equilibrium dynamics of closed many-body systems have attracted increasing interest from the scientific community over the past two decades, primarily due to the possibility of simulating model Hamiltonians in highly controllable quantum platforms, an approach known as analog quantum simulation. Prominent examples include cold atoms in optical lattices~\cite{greiner2002collapse, lewenstein2007ultracold, bloch2008many, gross2017quantum}, trapped ions~\cite{monroe2021programmable}, and Rydberg atom arrays~\cite{browaeys2020many}. In these platforms, system parameters can be fine tuned, and the system can be effectively isolated from the environment, allowing it to evolve coherently over long times. These features are difficult to achieve in solid-state experiments and are key ingredients for the protocol known as quantum quench: one prepares an initial state, typically the ground state of a given Hamiltonian, changes the Hamiltonian by modifying the system parameters, and lets the state evolve under the new Hamiltonian. More recently, digital quantum simulation on Noisy Intermediate-Scale Quantum (NISQ) devices has opened a complementary route to these problems~\cite{preskill2018quantum, ippoliti2021many}, although short coherence times still make it challenging to sustain unitary dynamics over long evolution times.

Sudden quenches, in which the system parameters are abruptly changed, have been extensively studied using analytical approaches~\cite{calabrese2006time} and numerical methods based on tensor networks~\cite{white2004real, schollwock2011density}. In analog simulators, they have been used to address fundamental questions such as light-cone effects that bound the speed of correlation propagation~\cite{lieb1972finite, calabrese2007quantum}, as well as equilibration and thermalization~\cite{eisert2015quantum, gogolin2016equilibration, polkovnikov2011colloquium}. However, sudden quenches across or close to quantum critical points, i.e., in the vicinity of a quantum phase transition (QPT), are much harder to probe. The final Hamiltonian has a vanishing gap and diverging relaxation times in this region~\cite{zurek2005dynamics}, so the system takes very long to equilibrate, a timescale that is hard to reach within the finite coherence times of current experiments. In addition, the abrupt protocol offers no control over how the critical region is crossed.

A natural way to gain this control is the slow quench, or ramp, in which a
system parameter is varied at a finite rate. We characterize the ramp by
the time $\tau$ spent per unit change of the parameter, so that $\tau^{-1}$
is the quench rate. The additional energy scale introduced by the ramp,
proportional to $\tau^{-1}$, can then be compared to the relaxation rates
and excitation gaps of the system, resulting in non-trivial
dynamics~\cite{schoenauer2019finite}.
For $\tau\to0$, the sudden quench is recovered, whereas for $\tau\to\infty$
the evolution is adiabatic and the system remains in the instantaneous
ground state. If the dynamics conserves a quantum number, however, the
evolved state is confined to the symmetry sector of the initial state, and
the adiabatic limit is the lowest-energy state within this sector, which
need not be the global ground state. The time required to reach this limit depends on the order and the universality class of the transition~\cite{carvalho2022formation}. For intermediate
$\tau$, the system loses adiabaticity close to the critical point, and
the final state contains a finite density of excitations. For
continuous transitions, this density follows a power law in $\tau$
determined by the critical exponents, according to the quantum
Kibble-Zurek mechanism~\cite{dziarmaga2010dynamics}. Across a BKT transition, where the gap
closes exponentially, this scaling acquires logarithmic corrections~\cite{pellegrini2008adiabatic}.

Different quantities can probe the dynamics of such quenches. The quantum work distribution~\cite{mascarenhas2014work, zawadzki2023non, tierz2025work} and the Loschmidt amplitude, whose zeros signal Dynamical Quantum Phase Transitions (DQPTs)~\cite{heyl2013dynamical, silva2008statistics, heyl2018dynamical, zhang2026fate}, are both determined by how the evolved state is distributed over the eigenstates of the Hamiltonian. When the state has a finite weight on several eigenstates, however, its dynamics is also encoded in the relative phases of these components, to which populations are blind. A quantity that captures these phases is the quantum-state texture recently introduced by Parisio~\cite{parisio2024quantum}, which measures how far the density matrix, written in a chosen basis, is from the uniform (textureless) one. 

Despite these advances, a systematic understanding of how the dynamics under
finite-time quenches depends on the order of the underlying QPT remains
incomplete. In this work, we address this question by analysing finite-time linear quenches on the
anisotropy of the XXZ Heisenberg model, for both spin-1/2 and spin-1 chains,
which at equilibrium undergo QPTs of different orders. This allows us to
compare directly how the quench-time dependence of the dynamics is affected
by the nature of the transition. Moreover, since the dynamics conserves the
total magnetization, ramps across the first-order transition probe its
manifestation at fixed magnetization, where the ferromagnetic phase is
replaced by phase separation into ferromagnetic domains. This situation is
directly relevant to quantum simulators in which the particle number is
conserved~\cite{bloch2008many, jepsen2020spin, hild2014far}.

Our analysis is based on exact diagonalization of chains with sizes
accessible to current analog and digital quantum simulators, which also
operate with a limited number of degrees of freedom. The full energy
spectrum allows us to resolve the decomposition of the evolved state in
the instantaneous eigenbasis and to identify the states populated during
the ramp. We discuss the behavior of local observables and of the
entanglement entropy, together with two quantities that depend only on
the populations of the instantaneous eigenstates, the fidelity with the
instantaneous ground state and the effective dimension, and with the
texture in the eigenbasis of the final Hamiltonian, which also depends on
their relative phases. We show that signatures of
the equilibrium QPTs are already imprinted on the dynamics of these small
systems.

This paper is organized as follows. In Sec.~\ref{sec:model}, we present
the model and its equilibrium properties, and describe the quench
protocol and the quantities used to characterize the dynamics.
Section~\ref{sec:results} presents the results for the spin-1/2 chain,
for ramps across the BKT and first-order transitions, and for the spin-1
chain, for ramps across the second-order transition. Section~\ref{sec:conclusions}
summarizes our conclusions.

\section{Model and methods}
\label{sec:model}
\subsection{Model and equilibrium phase diagram}

We consider anisotropic spin chains described by the XXZ Heisenberg
Hamiltonian
\begin{equation}
  H(t) = J \sum_{l=1}^{L-1} \left[ s^x_l s^x_{l+1} + s^y_l s^y_{l+1}
  + \Delta(t)\, s^z_l s^z_{l+1} \right] + h\, s^z_1 ,
  \label{eq:H}
\end{equation}
where $\mathbf{S}_l = (S^x_l, S^y_l, S^z_l)$ are spin-$S$ operators at
site $l$ of a chain with $L$ sites, with $S = 1/2$ or $1$, and
$\mathbf{s}_l = \mathbf{S}_l / S$ are normalized so that the eigenvalues
of $s^z_l$ lie in $[-1, 1]$; for $S = 1/2$, $\mathbf{s}_l =
\boldsymbol{\sigma}_l$ are the Pauli matrices. Open boundary conditions
are assumed. Here, $J$ is the transverse exchange coupling, $J\Delta$ is
the longitudinal one, and $\Delta$ is therefore the dimensionless
anisotropy. Throughout this paper, we set $J = \hbar = 1$, so that
energies are measured in units of $J$ and times in units of $\hbar/J$. 
The last term is a small longitudinal field on the first site, whose role
is discussed in Sec.~\ref{sec:protocol}.

\begin{figure}[t]
    \centering
    \includegraphics[width=\columnwidth]{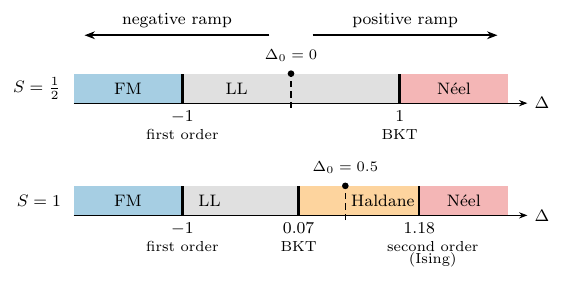}
   \caption{Ground-state phase diagrams of the spin-1/2 (top) and spin-1 (bottom) XXZ chains as a function of the anisotropy $\Delta$, with the order of each quantum phase transition indicated~\cite{giamarchi2003quantum, chen2003ground}. The dashed lines and arrows illustrate the protocol studied in this work: $\Delta$ is ramped linearly in time from an initial value $\Delta_0$, marked by the dashed lines (Sec.~\ref{sec:protocol}). For spin-1/2, the ramps start from $\Delta_0 = 0$, in the LL phase, and drive the system toward the N\'eel phase (positive ramp) or toward the FM phase (negative ramp). Since the dynamics conserves the total magnetization, the negative ramp leads to phase separation into ferromagnetic domains rather than to the FM ground state. For spin-1, the ramp starts from $\Delta_0 = 0.5$, in the Haldane phase, and crosses only the second-order Haldane--N\'eel transition.}
    \label{fig:phase_diagrams}
\end{figure}

At equilibrium, the spin-1/2 chain is integrable and can be solved by Bethe ansatz~\cite{giamarchi2003quantum}. When the longitudinal interaction dominates, the system displays magnetic order: a ferromagnetic (FM) phase for $\Delta \leq -1$ and a gapped antiferromagnetic (AFM) N\'eel phase for $\Delta> 1$. For $-1 < \Delta \leq 1$, the system is in a critical (gapless) Luttinger liquid (LL) phase. The two QPTs that bound the LL phase are of very different nature. The transition between the FM and LL phases, at $\Delta=-1$, is of first order, whereas the transition between the LL and N\'eel phases, at $\Delta=1$, is of infinite order and is known as a Berezinskii-Kosterlitz-Thouless (BKT) transition~\cite{kosterlitz1973ordering}. The BKT transition is not described by the Landau paradigm of symmetry breaking, and the gap of the N\'eel phase closes exponentially as $\Delta \to 1^+$~\cite{okamoto1996critical}.

The behavior of integer-spin chains is qualitatively different.
According to Haldane's conjecture, the isotropic antiferromagnetic chain with integer spin is gapped and has only short-range correlations~\cite{haldane1983continuum, haldane1983nonlinear, haldane2016ground}. The spin-1 XXZ model is not integrable, but its ground-state phase diagram is known~\cite{chen2003ground}. For negative $\Delta$, it resembles the spin-1/2 case: an FM phase for $\Delta \leq -1$ and a gapless LL phase for $-1 < \Delta \lesssim 0.07$, separated by a first-order QPT. For $0.07 \lesssim \Delta \lesssim 1.18$, the system is in the Haldane phase, a gapped symmetry-protected topological phase~\cite{pollmann2012detection, kennedy1992hidden}, and the LL--Haldane transition is also of the BKT type. Finally, for $\Delta \gtrsim 1.18$, the system enters the N\'eel phase, and the Haldane--N\'eel transition is of second order, in the Ising universality class. Thus, the two models together provide QPTs of first, second, and infinite order (Fig.~\ref{fig:phase_diagrams}).

\subsection{Quench protocol}
\label{sec:protocol}

We start from the ground state at $\Delta_0$ and vary the anisotropy
linearly in time, so that it reaches a target value $\Delta_f$ at
$t=t_f$,
\begin{equation}
\Delta(t) = \Delta_0 + {\rm sgn}(\Delta_f-\Delta_0)\,\frac{t}{\tau},
\label{eq:ramp}
\end{equation}
with $t_f=|\Delta_f-\Delta_0|\,\tau$. Hence, $\tau$ is the time spent per
unit change of $\Delta$, i.e., the inverse of the quench rate.
For the spin-1/2 chain, we take $\Delta_0=0$, in
the LL phase, and $\Delta_f=\pm5$, so that the positive and negative ramps
cross the BKT and first-order transitions, respectively. For the spin-1
chain, we focus on the second-order transition and ramp from $\Delta_0=0.5$,
inside the Haldane phase, to $\Delta_f=3$, crossing only the Haldane--N\'eel
transition.

\begin{figure}[t]
    \centering
    \includegraphics[width=\columnwidth]{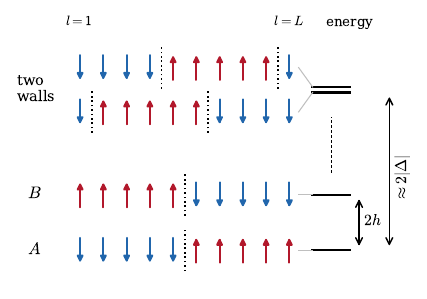}
   \caption{Low-energy structure of the $S^z_{\rm tot}=0$ sector of the
spin-$1/2$ chain deep in the ferromagnetic regime, $|\Delta|\gg1$, with
each configuration drawn at the height of its energy level (not to scale).
The single-wall states $A$ and $B$ are related by a global spin flip and
split by the pinning field. The
two lowest two-wall states are related by spatial reflection and nearly
degenerate. Dotted lines mark the domain walls.}
    \label{fig:esquema_paredes}
\end{figure}

The Hamiltonian conserves the total magnetization,
$[H(t),S^z_{\rm tot}]=0$, with $S^z_{\rm tot}=\sum_l S^z_l$. Since the
initial state has $S^z_{\rm tot}=0$, the evolved state remains in this
sector at all times, and the natural reference for the dynamics is the
lowest-energy state within the sector. For $\Delta>-1$, this state coincides
with the global ground state. For $\Delta<-1$, in contrast, the global ground
state is ferromagnetic and lies in a different sector, while the
lowest-energy state with $S^z_{\rm tot}=0$ is phase separated into
ferromagnetic domains of opposite orientation.
The structure of the $S^z_{\rm tot}=0$ sector in this regime is simple in
the Ising limit, $|\Delta|\gg1$, in which the transverse terms of
Eq.~\eqref{eq:H} can be neglected, Fig.~\ref{fig:esquema_paredes}.
Each antiparallel bond then costs an energy $2|\Delta|$ relative to a
parallel one, so that, with open boundaries, the lowest-energy
configurations contain a single domain wall at the center of the chain. The
two such configurations, $A$ and $B$, are related by a global spin flip. 
Configurations with two walls lie higher in energy by $\approx2|\Delta|$. 
Domain-wall (kink) ground states of the ferromagnetic XXZ chain at
fixed magnetization are well known~\cite{alcaraz1995anisotropic, koma1997spectral}; here the number of domain walls provides a
natural measure of the excitations created by the ramp. The transverse
terms exchange neighboring antiparallel spins and therefore move the walls
by one site; for $|\Delta|$ well above unity, they only dress these
configurations, with energy corrections of order $1/|\Delta|$, and do not
change this classification.

In both ordered regimes, the lowest-energy states of the $S^z_{\rm tot}=0$
sector are quasi-degenerate: the two N\'eel configurations for large
positive $\Delta$, and the single-wall states $A$ and $B$ for $\Delta<-1$.
The pinning field $h\,s^z_1$ in Eq.~(\ref{eq:H}), with $h=0.1$ unless
stated otherwise, selects one of them; for $\Delta<-1$, it splits $A$ and
$B$ by $2h$ (Fig.~\ref{fig:esquema_paredes}). Since it preserves
$S^z_{\rm tot}$ and acts on a single site, it does not affect the bulk
phase diagram discussed above.

The evolved state is 
$U(t)|\psi(0)\rangle$, where $|\psi(0)\rangle$ is the ground state of $H(0)$, and 
$U(t)=\mathcal{T}\exp\left[-i\int_0^t dt'\,H(t')\right]$ is the time-evolution
operator, with $\mathcal{T}$ the time-ordering operator. In practice, we
compute $U(t)$ by approximating $H(t)$ as constant within small time steps
$\delta t$,
\begin{equation}
U(t)\simeq\prod_{k}\exp\left[-i\,H(t_k)\,\delta t\right],
\end{equation}
where each factor is evaluated exactly by diagonalizing $H(t_k)$. The time
step is chosen proportional to the ramp time, $\delta t=\delta\Delta\,\tau$,
so that the anisotropy changes by the same amount $\delta\Delta=0.01$
in each step, independently of $\tau$. We verified that halving
$\delta\Delta$ does not change the results within the precision of the
figures.

\subsection{Characterization of the dynamics}
\label{sec:characterization}

A natural way to characterize the phases reached by the evolved state is
through its total and staggered magnetizations per site,
\begin{equation}
  M(t) = \frac{1}{L} \sum_{l=1}^{L} \langle s^z_l \rangle_t, \qquad
  M_s(t) = \frac{1}{L} \sum_{l=1}^{L} (-1)^l \langle s^z_l \rangle_t ,
  \label{eq:magnetizations}
\end{equation}
where $\langle\cdot\rangle_t$ denotes the expectation value in $|\psi(t)\rangle$, and through the entanglement entropy between the two halves of the chain,
\begin{equation}
EE(t)=-\mathrm{Tr}\left[\rho_{L/2}(t)\log_2\rho_{L/2}(t)\right],
\label{eq:EE}
\end{equation}
where $\rho_{L/2}(t)=\mathrm{Tr}_{l>L/2}\,|\psi(t)\rangle\langle\psi(t)|$ is the reduced density matrix of the first $L/2$ sites, obtained by tracing out the remaining ones.

To assess how far from equilibrium the dynamics is for each $\tau$, we
compare these quantities with their instantaneous ground-state values,
obtained by replacing $|\psi(t)\rangle$ with the instantaneous ground state.

The extent to which the Hilbert space is explored during the evolution is
encoded in the transition probabilities
\begin{equation}
p_m(t) = |\langle\psi_m^t|\psi(t)\rangle|^2,
\label{eq:pm}
\end{equation}
where $|\psi_m^t\rangle$ is the $m$th eigenstate of the instantaneous
Hamiltonian $H(t)$, with energy $E_m^t$, ordered by increasing energy. In
particular, $f(t)\equiv p_0(t)$ is the fidelity between the evolved state
and the instantaneous ground state. Since $U(t)$ preserves
$S^z_{\rm tot}$, $p_m(t)=0$ for every eigenstate outside the
$S^z_{\rm tot}=0$ sector, and we denote by $f_0(t)$ the fidelity with the
lowest-energy eigenstate within this sector, which coincides with $f(t)$
for $\Delta>-1$. 
To compare different ramps, we adopt the criterion of
Ref.~\cite{carvalho2022formation} and consider the evolution essentially
adiabatic when the fidelity at the end of the ramp exceeds $0.99$, with
$f$ replaced by $f_0$ for the negative ramp. We denote by $\tau_{\rm ad}$
the shortest ramp time that satisfies this criterion.

The transition probabilities also define the effective
dimension of the evolved state in the instantaneous eigenbasis~\cite{dos2024smooth},
\begin{equation}
d_{\rm eff}(t) = \frac{1}{\sum_m p_m^2(t)},
\label{eq:deff}
\end{equation}
the inverse of the inverse participation ratio (IPR), explored in the
context of quantum chaos~\cite{borgonovi2016quantum}. It ranges from
$d_{\rm eff}=1$, when a single instantaneous eigenstate is occupied, to
$d_{\rm eff}=D$, the dimension of the $S^z_{\rm tot}=0$ sector, when all
its eigenstates are equally populated.

\subsection{Quantum-state texture}
\label{sec:texture}

The fidelity and the effective dimension depend only on the populations
$p_m(t)$ and are insensitive to the relative phases of the occupied
eigenstates. To access these phases, we use the resource theory of
quantum-state texture introduced by Parisio~\cite{parisio2024quantum},
which has since been used to probe equilibrium
QPTs~\cite{Casagrande2026}. Given an orthonormal basis $\{|k\rangle\}$ of
a $D$-dimensional space, the only state whose density matrix has all
entries equal is the textureless state
$|f_1\rangle = D^{-1/2}\sum_k |k\rangle$. The texture of a state $\rho$ is
quantified by its rugosity~\cite{parisio2024quantum},
\begin{equation}
  \begin{aligned}
    \gsum(\rho) &= D\,\langle f_1|\rho|f_1\rangle = \sum_{k,l}\rho_{kl},\\
    \mathcal{R}(\rho) &= -\ln\langle f_1|\rho|f_1\rangle
    = -\ln\frac{\gsum(\rho)}{D},
  \end{aligned}
  \label{eq:texture}
\end{equation}
where $\gsum(\rho)$ is the grand sum of $\rho$, i.e., the sum of all its
entries in the chosen basis. The rugosity vanishes only for the
textureless state, and $\gsum(\rho)/D$ is the probability of finding the
state in $|f_1\rangle$, so that $0\leq\gsum(\rho)\leq D$. Since
$\mathcal{R}$ and $\gsum$ carry the same information, we show $\gsum$,
whose deviations from unity are directly the interference terms discussed
below.

In previous applications to QPTs~\cite{Casagrande2026}, the reference
basis is the computational one. Then $|f_1\rangle$ is a product state,
$\gsum(\rho)$ is a sum of transverse spin correlators of all orders, and
$\mathcal{R}$ bounds the geometric entanglement from
above~\cite{Casagrande2026}. Here, we take instead as reference basis the
eigenstates $\{|\psi^f_m\rangle\}$ of the final Hamiltonian $H(t_f)$
within the $S^z_{\rm tot}=0$ sector, so that $D$ is the dimension of this
sector. In this basis, $|f_1\rangle$ is not a product state, and the grand
sum probes instead the relative phases of energy eigenstates. For the
evolved pure state, with $c_m(t) = \langle\psi^f_m|\psi(t)\rangle$,
\begin{equation}
  \gsum(t) = \Big|\sum_m c_m(t)\Big|^2
  = 1 + \sum_{m\neq n} c_m(t)\, c_n^*(t).
  \label{eq:texture_t}
\end{equation}
The diagonal terms always add to unity, and all the information is in the
interference terms: a single eigenstate of $H(t_f)$ gives $\gsum(t) = 1$,
whereas a superposition gives $\gsum(t)$ above or below unity, depending
on the relative phases.

The grand sum is computed in this fixed basis, whereas the populations
$p_m(t)$ refer to the instantaneous eigenbasis. The two bases coincide at
$t = t_f$, where $|c_m(t_f)|^2 = p_m(t_f)$, and the triangle inequality
gives $\gsum(t_f) \le \big[\sum_m \sqrt{p_m(t_f)}\big]^2$, with equality
when all $c_m(t_f)$ have the same phase. At earlier times,
$|c_m(t)|^2 \neq p_m(t)$ in general, but the two are close once the
instantaneous eigenstates approach those of $H(t_f)$, which happens
shortly after the transitions considered here. The grand sum is thus
complementary to $d_{\rm eff}$: both are trivial when a single eigenstate
is occupied, but only $\gsum(t)$ follows the relative phases of a
superposition. We illustrate this for the negative ramp in
Sec.~\ref{subsec:rampa_negativa}.

\section{Results}
\label{sec:results}

\subsection{Spin-1/2 chain}

\subsubsection{Positive ramp}
\label{subsec:rampa_positiva}

We first consider ramps toward the N\'eel phase, starting from the ground
state at $\Delta_0 = 0$ and crossing the BKT transition at $\Delta = 1$.
Before discussing the dynamics, we examine how signatures of this
transition manifest themselves in observables of finite chains. In
contrast to the first-order transition, which remains sharp for any chain
length (Sec.~\ref{subsec:rampa_negativa}), the BKT transition has no
sharp signature: the staggered
magnetization of the ground state grows only well above $\Delta = 1$. We
therefore locate the point at which the finite chain develops N\'eel order
by the maximum of $dM_s/d\Delta$, which defines a pseudo-critical point
$\Delta^*(L)$ (Appendix~\ref{app:pseudocritical}), with $\Delta^*\approx3.4$
for $L=10$. Since $\Delta^*(L)$ also depends on the pinning field, it is
not an estimate of the transition point; as we show next, however, it is
the relevant scale for the ramp dynamics.
Throughout this paper, “transition” refers to the thermodynamic limit; in the finite chain, the BKT transition appears only as a smooth crossover around $\Delta^*$.

\begin{figure}[t]
    \centering
    \includegraphics[width=\columnwidth]{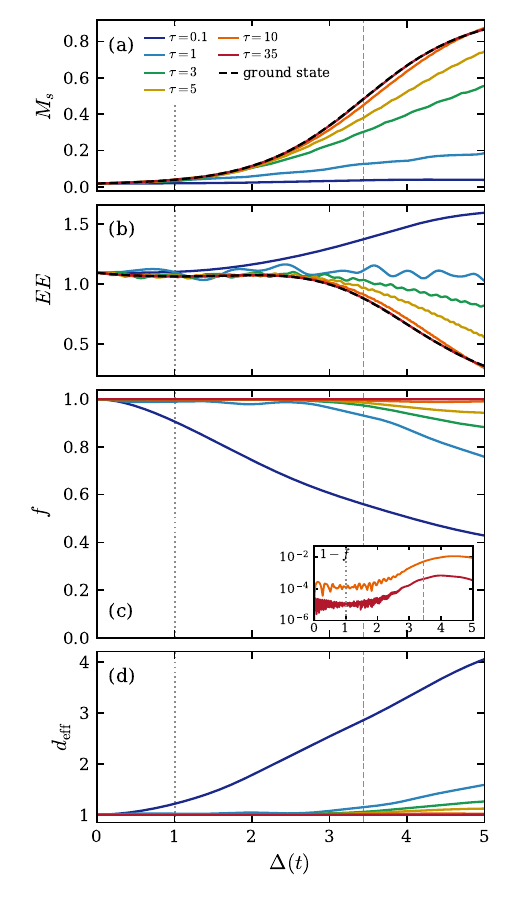}
    \caption{Positive ramp of the spin-$1/2$ chain with $L=10$, from
$\Delta_0=0$ to $\Delta_f=5$, for $\tau=0.1$, $1$, $3$, $5$, $10$, and $35$:
(a) staggered magnetization $M_s$, (b) entanglement entropy $EE$,
(c) fidelity $f$ with the instantaneous ground state, and (d) effective
dimension $d_{\rm eff}$, as functions of $\Delta(t)$. The inset in (c) shows $1-f$ for $\tau=10$ and $35$ on a logarithmic scale. Dashed black lines in (a) and
(b) show the instantaneous ground-state values. The dotted lines mark the
BKT transition at $\Delta=1$, and the dashed gray lines mark the pseudo-critical point
$\Delta^* \approx 3.4$ of the $L=10$ chain (Appendix~\ref{app:pseudocritical}).}
    \label{fig:rampa_positiva}
\end{figure}

Figure~\ref{fig:rampa_positiva} shows the dynamics of the quantities
defined in Sec.~\ref{sec:characterization} for $L=10$. The
staggered magnetization, Fig.~\ref{fig:rampa_positiva}(a), signals the
development of N\'eel order. For the fastest ramp, $\tau=0.1$, the quench
is nearly sudden and $M_s$ remains close to its initial value: the state
has no time to adapt to the changing Hamiltonian, and N\'eel order does not
develop. As $\tau$ increases, $M_s$ approaches the ground-state curve and,
for $\tau\gtrsim10$, follows it over the whole ramp. The entanglement
entropy, Fig.~\ref{fig:rampa_positiva}(b), shows the same trend. In the
ground state, $EE$ decreases as $\Delta$ grows, since the N\'eel state
approaches a product state, whereas for the fastest ramp it grows above its
initial value, as the excitations created by the ramp spread entanglement
across the chain. Intermediate ramps lie between these two limits, with
oscillations that reflect the interference between the occupied
eigenstates. The fidelity and the effective dimension,
Figs.~\ref{fig:rampa_positiva}(c) and~\ref{fig:rampa_positiva}(d), show
that this agreement reflects genuine ground-state tracking rather than a
coincidence of average values: for $\tau\gtrsim10$, $f$ remains close to
unity and $d_{\rm eff}\simeq1$ throughout the evolution.

Although the BKT transition occurs at $\Delta=1$ in the thermodynamic
limit, the fidelity is not lost there for $L=10$. According to the
adiabatic theorem, the evolved state follows the instantaneous ground
state as long as the latter changes slowly on the scale set by the
gap~\cite{de2010adiabatic}. The fidelity is therefore lost where the
ground state changes rapidly, which happens as it develops N\'eel order
around $\Delta^*$ (Appendix~\ref{app:pseudocritical}). Accordingly, for
intermediate ramps, $1\leq\tau\leq5$, $f$ remains close to unity well
beyond $\Delta=1$ and decreases only as the ramp approaches $\Delta^*$,
Fig.~\ref{fig:rampa_positiva}(c). For slow ramps, the loss of fidelity is
too small to be seen on the scale of the main panel, and the inset shows
$1-f$ for $\tau=10$ and $35$ on a logarithmic scale. In both cases, $1-f$
grows by more than an order of magnitude across $\Delta^*$, reaches a
maximum after it, and decreases again
deep in the N\'eel phase, where the ground state changes slowly. Thus, for the positive ramp, moderate ramp times suffice for an
essentially adiabatic evolution of the finite chain, with
$\tau_{\rm ad}\approx10$, and the non-adiabatic effects are associated
with the pseudo-critical point rather than with the BKT transition
itself. As shown next, the negative ramp behaves very
differently.

\subsubsection{Negative ramp}
\label{subsec:rampa_negativa}

\begin{figure}[t]
    \centering
    \includegraphics[width=\columnwidth]{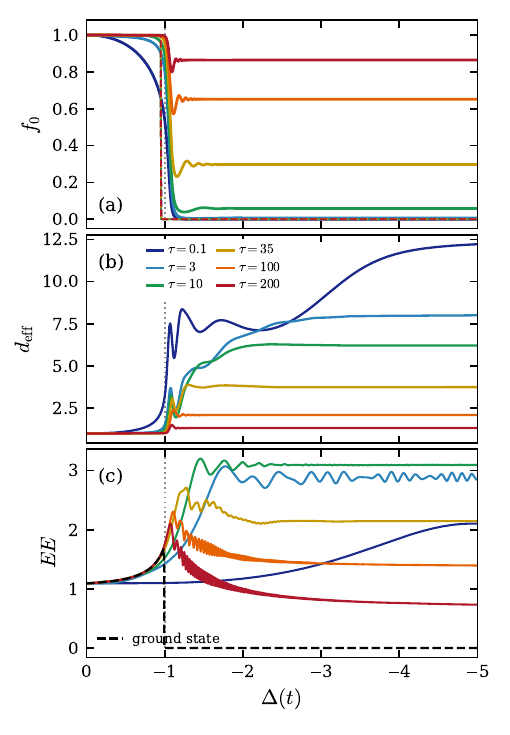}
    \caption{Negative ramp of the spin-$1/2$ chain with $L=10$, from
$\Delta_0=0$ to $\Delta=-5$, for $\tau=0.1$, $3$, $10$, $35$, $100$, and
$200$: (a) fidelity $f_0$ with the lowest-energy eigenstate of the
$S^z_{\rm tot}=0$ sector (solid lines) and fidelity $f$ with the global
ground state (dashed lines), (b) effective dimension $d_{\rm eff}$, and
(c) entanglement entropy $EE$, as functions of $\Delta(t)$. The dashed black
line in (c) shows the ground-state value, and the dotted lines mark
$\Delta=-1$.}
    \label{fig:rampa_negativa}
\end{figure}

The negative ramp crosses the first-order transition at $\Delta=-1$. This
transition is a level crossing and therefore remains sharp for any chain
length: at $\Delta=-1$, the ground state changes from the
$S^z_{\rm tot}=0$ ground state of the LL phase to a fully polarized
ferromagnetic state, with $M=-1$. Since the
dynamics conserves $S^z_{\rm tot}$, the evolved state cannot follow this
level crossing: its magnetization remains zero, and its overlap with the
global ground state, which lies in a different sector, vanishes by
symmetry for $\Delta<-1$. The fidelity $f$ thus vanishes as soon as the ramp crosses $\Delta=-1$ for all ramp times, dashed lines in Fig.~\ref{fig:rampa_negativa}(a), which
simply marks the level crossing. Adiabaticity must instead be assessed
with respect to the lowest-energy state within the sector, the single-wall
state $A$ (Fig.~\ref{fig:esquema_paredes}), through $f_0$.

For $\tau\geq3$, $f_0$ remains close to unity in the LL phase, decreases while $\Delta$ goes from $-1$ to about $-1.5$, and remains constant afterwards, at a value that approaches unity as $\tau$ increases, Fig.~\ref{fig:rampa_negativa}(a). The effective dimension,
Fig.~\ref{fig:rampa_negativa}(b), grows at the crossing and then settles
on a plateau, which decreases with $\tau$ from several occupied
eigenstates to essentially one. The populations of the instantaneous
eigenstates, and with them the number of domain walls, are thus fixed
shortly after the ramp crosses $\Delta=-1$. The fastest ramp, $\tau=0.1$,
is the exception: the quench is nearly sudden, $f_0$ decreases already in
the LL phase, 
and $d_{\rm eff}$ keeps growing for all $\Delta < - 1$, since the evolved state remains close to the initial one while the instantaneous eigenbasis keeps changing. 
Slower ramps leave fewer excitations, but the approach to the
single-wall state is slow. As shown in Appendix~\ref{app:finite_size},
the weight outside $A$ at the end of the ramp decays as
$1-f_0(t_f)\simeq e^{-\tau/\tau_0}$, the form expected for a
Landau--Zener passage through an avoided crossing~\cite{zener1932,de2010adiabatic},
with $\tau_0\approx 100$ for $L=10$ [Fig.~\ref{fig:finite_size}(b)].
Extrapolating this fit to the criterion of Sec.~\ref{sec:characterization},
$1-f_0(t_f)=10^{-2}$, gives $\tau_{\rm ad}\approx 460$, so that an
essentially adiabatic evolution requires ramps more than an order of
magnitude slower than in the positive ramp, for which
$\tau_{\rm ad}\approx 10$. This separation of timescales grows with the
chain length, since $\tau_0$ roughly doubles every two sites, whereas
$\tau_{\rm ad}$ of the positive ramp grows only slowly.
The entanglement entropy,
Fig.~\ref{fig:rampa_negativa}(c), also settles on plateaus that decrease
with $\tau$ for slow ramps; since the single-wall state is nearly a product
state across the central cut, $EE$ is particularly sensitive to the residual
components. For the fastest ramps, in contrast, $EE$ grows slowly, because
the ramp reaches $\Delta_f$ before the entanglement has time to spread
across the chain.

The populations fixed after the ramp crosses $\Delta=-1$ separate two
effects. Repeating the ramp for different pinning fields
(Appendix~\ref{app:pinning}), we find that the fraction of the population
in single-wall states is nearly independent of $h$, so that the number of
domain walls created by the ramp is an intrinsic property of the dynamics
across the first-order transition. The distribution between $A$ and $B$,
in contrast, is controlled by their splitting $2h$, and approaches equal
weights as $h\to0$.

\begin{figure*}[t]
    \centering
    \includegraphics[width=\textwidth]{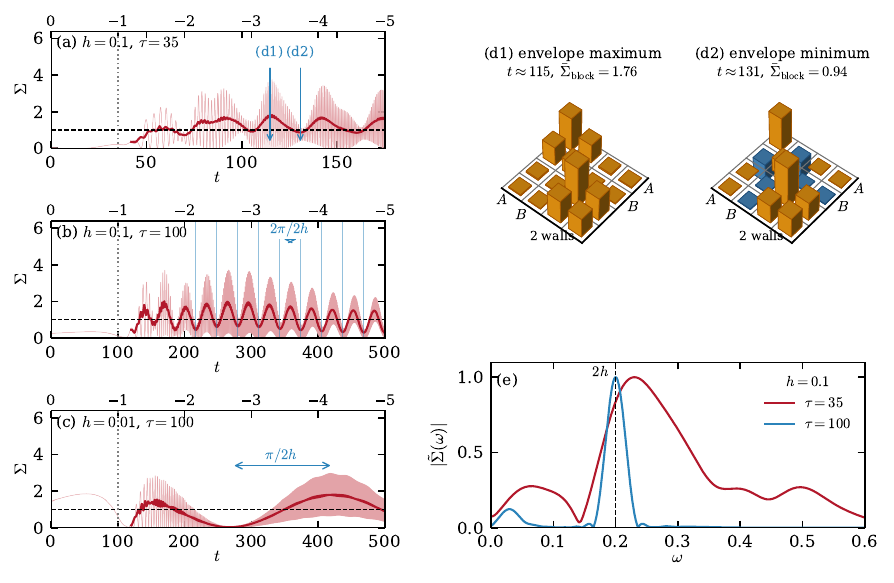}
    \caption{Grand sum of the evolved state during the negative ramp of the
spin-$1/2$ chain with $L=10$, in the eigenbasis of $H(t_f)$ within the
$S^z_{\rm tot}=0$ sector ($D=252$).
(a)--(c) $\gsum(t)$ as a function of $t$ (bottom axes) and $\Delta(t)$
(top axes) for (a) $h=0.1$, $\tau=35$, (b) $h=0.1$, $\tau=100$, and
(c) $h=0.01$, $\tau=100$. Light lines show $\gsum(t)$, and dark lines its
average over a sliding window of width $8$; the dashed lines mark
$\gsum=1$, the value for a single eigenstate of $H(t_f)$, and the dotted
lines mark $\Delta=-1$. The arrows in (a) mark the times shown in (d1)
and (d2), the vertical lines in (b) are separated by $2\pi/2h$, and the
arrow in (c) spans $\pi/2h$.
(d1), (d2) Real part of the density matrix, averaged over the same
window, in the block of the four most populated eigenstates of $H(t_f)$
(the single-wall states $A$ and $B$ and two two-wall states), at
(d1) a maximum and (d2) the following minimum of the envelope of
$\gsum(t)$ for $h=0.1$ and $\tau=35$, in the bar-chart representation of
Ref.~\cite{parisio2024quantum}; orange (blue) bars are positive
(negative), and $\bar\gsum_{\rm block}$ is the grand sum of this block.
(e) Fourier spectrum of $\gsum(t)$ for $t>1.5\tau$ and $h=0.1$,
normalized to its maximum; the dashed line marks $\omega=2h$.}
    \label{fig:texture}
\end{figure*}

Once the populations are frozen, $f_0$ and $d_{\rm eff}$ no longer
evolve, but the state is a coherent superposition of domain-wall states
whose relative phases keep evolving. This dynamics is captured by the
grand sum in the eigenbasis of $H(t_f)$,
Figs.~\ref{fig:texture}(a)--\ref{fig:texture}(c). Before the crossing
[dotted lines in Figs.~\ref{fig:texture}(a)--\ref{fig:texture}(c)],
$\gsum(t) \ll 1$, since the evolved state has a small overlap with the
textureless state of the final basis. As the population is transferred
to the domain-wall states, $\gsum(t)$ rises to values of order unity and
then oscillates on two time scales: fast oscillations, whose frequency
increases along the ramp, modulated by a slow envelope.

To identify the origin of these two time scales, we separate the
amplitudes $c_A$ and $c_B$ of the single-wall states
(Fig.~\ref{fig:esquema_paredes}) from the sum $c_X = \sum_{m\in X} c_m$
over the states with two or more walls. Equation~\eqref{eq:texture_t}
then reads
\begin{equation}
  \gsum(t) = |c_A + c_B|^2 + 2\,\mathrm{Re}\big[(c_A + c_B)\,c_X^*\big]
  + |c_X|^2 .
  \label{eq:texture_ABX}
\end{equation}
After the crossing, the instantaneous eigenstates are close to those of
$H(t_f)$, and each component only accumulates a dynamical phase,
\begin{equation}
  c_m(t) \simeq |c_m|\,
  \exp\!\Big[i\varphi_m - i\int_{t^*}^{t} E_m^{t'}\,dt'\Big],
  \label{eq:dynamical_phase}
\end{equation}
where $t^*$ is any time after the populations freeze and the constant
$\varphi_m$ includes the phase acquired across the transition. The
relative phases are therefore set by energy differences.
A two-wall state has one more antiparallel bond than a single-wall
state and thus lies higher in energy by $\approx 2|\Delta|$
(Fig.~\ref{fig:esquema_paredes}). The relative phase between the single-
and two-wall components, which enters the second term of
Eq.~\eqref{eq:texture_ABX}, therefore advances at the instantaneous
rate $2|\Delta(t)|$. This term produces the fast oscillations, whose
frequency increases along the ramp. The single-wall states $A$ and $B$,
in contrast, are split only by the pinning field, by $2h$, and their
relative phase advances at the much slower rate $2h$. This slow phase
enters the first term of
Eq.~\eqref{eq:texture_ABX},
\begin{equation}
  |c_A + c_B|^2 = |c_A|^2 + |c_B|^2
  + 2|c_A||c_B|\cos\!\big[\varphi_{AB} - 2h(t - t^*)\big],
  \label{eq:envelope}
\end{equation}
with $\varphi_{AB} = \varphi_A - \varphi_B$. It therefore modulates both
the mean value of $\gsum(t)$, through Eq.~\eqref{eq:envelope}, and the
amplitude of the fast oscillations, which is proportional to
$|c_A + c_B|$, giving rise to the slow envelope.

The envelope is thus a coherence effect between $A$ and $B$, which can
be seen directly in the density matrix. Figures~\ref{fig:texture}(d1)
and \ref{fig:texture}(d2) show its block in the four most populated
eigenstates of $H(t_f)$, at a maximum and at the following minimum of
the envelope for $\tau=35$ [arrows in Fig.~\ref{fig:texture}(a)],
averaged over a window long compared with the period of the fast
oscillations but short compared with that of the envelope. The
populations, given by the diagonal bars, are the same in both panels.
The off-diagonal bars that connect single- and two-wall states are
nearly flat, since these coherences oscillate at the fast frequency and
average out. What changes between the two panels is the pair of
off-diagonal bars for the coherence $\rho_{AB} = c_A c_B^*$, whose real
part gives the last term of Eq.~\eqref{eq:envelope}: they are positive
(orange) at the maximum of the envelope and negative (blue) at the
minimum.

The time scale of the envelope is set by the pinning field. For
$h=0.1$, the minima of the envelope are separated by $2\pi/2h$,
Fig.~\ref{fig:texture}(b), and the Fourier spectrum of $\gsum(t)$ after
the transition is peaked close to $\omega = 2h$,
Fig.~\ref{fig:texture}(e); the peak is broader for $\tau=35$, since
the evolution after the transition is shorter. Reducing the field to $h = 0.01$
slows down the envelope by an order of magnitude, while the fast
oscillations are unchanged, Fig.~\ref{fig:texture}(c).
The grand sum thus exposes the coherence between the two single-wall
configurations, which is invisible to $f_0$ and $d_{\rm eff}$.

In summary, at fixed magnetization the first-order transition manifests
itself dynamically as phase separation into ferromagnetic domains.

\subsection{Spin-1 chain}
\label{subsec:spin1}

The spin-1 chain allows us to include a second-order transition in the
comparison. To isolate it, we start from $\Delta_0=0.5$, inside the Haldane
phase, so that the positive ramp crosses only the Haldane--N\'eel
transition, in the Ising universality class, at $\Delta\approx1.18$. We
consider a chain with $L=6$, whose Hilbert-space dimension, $3^6=729$, is
comparable to that of the spin-1/2 chain with $L=10$. Negative ramps, which
cross the first-order transition at $\Delta=-1$, behave as in the spin-1/2
chain (not shown), reflecting the phase separation imposed by the
conservation of $S^z_{\rm tot}$.

\begin{figure}[!tbp]
    \centering
    \includegraphics[width=\columnwidth]{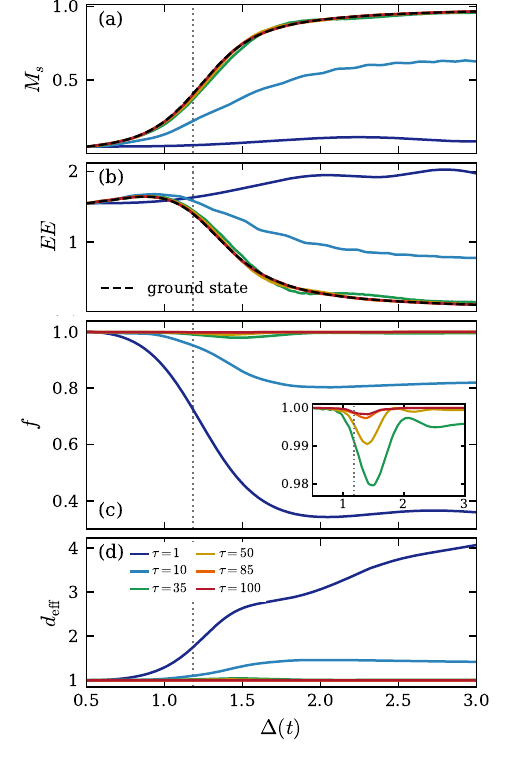}
    \caption{Positive ramp of the spin-1 chain with $L=6$, from
$\Delta_0=0.5$ to $\Delta_f=3$, for $\tau=1$, $10$, $35$, $50$, $85$,
and $100$: (a) staggered magnetization $M_s$, (b) entanglement entropy
$EE$, (c) fidelity $f$ with the instantaneous ground state, and
(d) effective dimension $d_{\rm eff}$, as functions of $\Delta(t)$.
Dashed black lines in (a) and (b) show the instantaneous ground-state
values. The inset in (c) shows $f$ for $\tau\geq35$ on an expanded scale.
The dotted vertical lines mark the Haldane--N\'eel transition in the
thermodynamic limit, $\Delta\approx1.18$.}
    \label{fig:spin1_positiva}
\end{figure}

Figure~\ref{fig:spin1_positiva} shows $M_s$, $EE$, $f$, and
$d_{\rm eff}$ during the positive ramp. As for spin-1/2, slowing down the
ramp leads to ground-state tracking: for fast ramps, $M_s$ remains close
to its initial value, $EE$ does not decrease as in the ground state, $f$
decreases, and $d_{\rm eff}$ grows, whereas for $\tau\gtrsim35$, $M_s$
and $EE$ follow their ground-state values, $f$ remains close to unity
throughout the ramp, and $d_{\rm eff}\simeq1$.
The loss of fidelity is concentrated in a narrow window around the
transition, Fig.~\ref{fig:spin1_positiva}(c). For fast ramps, $f$ then
remains essentially constant, whereas for slow ramps it recovers, as
discussed below. The second-order transition thus leaves a localized
dynamical signature already for $L=6$.

For slow ramps, $f$ dips near the transition and then recovers deeper in
the N\'eel phase, to values that approach unity as $\tau$ increases
[inset of Fig.~\ref{fig:spin1_positiva}(c)]. The
mechanism is the same as for the spin-$1/2$ chain
(Sec.~\ref{subsec:rampa_positiva}): the fidelity is lost where the
instantaneous ground state changes rapidly. What differs is where this
happens, and this difference reflects the order of the transition. At this
second-order transition, the gap closes as a power law, and already for
$L=6$ the ground state changes rapidly only in a narrow window close to
the thermodynamic-limit transition, where the loss of fidelity is
concentrated. At the BKT transition, in contrast, the gap closes
exponentially, finite-size effects are much stronger, and the region where
the ground state changes rapidly lies far from $\Delta=1$ at these system
sizes.

\begin{figure}[!tbp]
    \centering
    \includegraphics[width=\columnwidth]{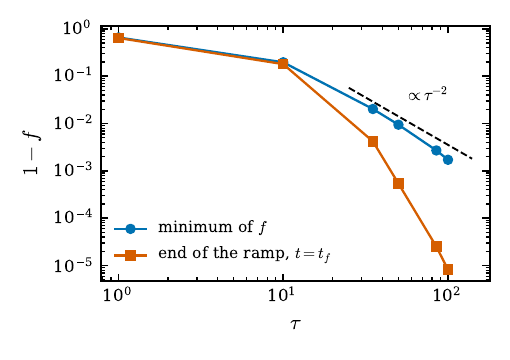}
    \caption{$1-f$ as a function of $\tau$ for the positive ramp of the
    spin-1 chain with $L=6$, evaluated at the minimum of the fidelity during
    the ramp, $f_{\min}$, and at the end of the ramp, $t=t_f$. The dashed
    line is proportional to $\tau^{-2}$, the dependence expected from
    adiabatic perturbation theory for the crossing of a finite
    gap~\cite{de2010adiabatic}.}
    \label{fig:spin1_slow}
\end{figure}

Figure~\ref{fig:spin1_slow} quantifies this behavior. For slow ramps, the
maximum loss of fidelity during the ramp, $1-f_{\rm min}$, decays
approximately as $\tau^{-2}$, the dependence expected from adiabatic
perturbation theory when the ramp crosses a finite
gap~\cite{de2010adiabatic}. We should stress that, for this
chain length, the minimum gap at the transition remains finite on
the scale set by the ramp. The loss of fidelity at the end of the ramp
decreases much faster: for slow ramps, almost all of the population lost
near the transition returns to the ground state once the gap reopens.

\section{Conclusions}
\label{sec:conclusions}

We have studied finite-time linear ramps of the anisotropy in spin-1/2 and
spin-1 XXZ chains, which together host quantum phase transitions of first,
second, and infinite order. Even at the small sizes accessible to exact
diagonalization, the order of the transition determines how adiabaticity is
lost and recovered. 
Across the BKT transition of the spin-1/2 chain, ground-state tracking is
recovered for moderate ramp times, and the loss of fidelity is associated
with a pseudo-critical point where the finite chain develops N\'eel order,
far from the thermodynamic-limit transition at $\Delta=1$ and approaching
it only slowly with the system size.
Across the second-order Haldane--N\'eel transition of the spin-1 chain,
in contrast, the loss of fidelity is concentrated in a narrow window close
to the thermodynamic-limit transition already for $L=6$, and is largely
recovered once the gap reopens.
Across the first-order transition, the conservation
of the total magnetization turns the ferromagnetic phase into phase
separation: slow ramps form a single domain wall, faster ramps create
additional walls, and the number of walls is fixed shortly after the
transition, whereas concentrating the population in the lowest single-wall state
requires ramps more than an order of magnitude slower than across the BKT
and second-order transitions, a separation that grows with the system size (Appendix~\ref{app:finite_size}).

Once the populations freeze, the remaining dynamics lies entirely in the
relative phases of the occupied eigenstates, to which the fidelity and the
effective dimension are blind. 
The grand sum $\gsum$ in the eigenbasis of the final Hamiltonian resolves
this dynamics as a beating between domain-wall
states, with a fast frequency set by the cost of a domain wall and an
envelope set by the field splitting. 
Since $\gsum/D$ is the fidelity with a single reference state, the
textureless state, it provides a simple probe of this coherence.

These results are directly relevant to quantum simulators, which operate
with a comparable number of degrees of freedom and often conserve the
particle number, the analog of the total magnetization. In this setting, a
ramp across a first-order transition does not prepare the symmetry-broken
phase, but a phase-separated state, and the number of domain walls provides
a natural measure of the excitations created by the ramp. Extending this
analysis to larger systems with the density matrix renormalization group
will allow a systematic finite-size scaling, in particular a test of
Kibble--Zurek scaling across the second-order transition and of the
statistics of domain-wall formation across the first-order one.

\begin{acknowledgments}
This work was supported by the Conselho Nacional de Desenvolvimento
Cient\'{\i}fico e Tecnol\'ogico (CNPq), the Coordena\c{c}\~ao de
Aperfei\c{c}oamento de Pessoal de N\'{\i}vel Superior (CAPES), the
Funda\c{c}\~ao de Apoio \`a Pesquisa do Estado de Minas Gerais (FAPEMIG), 
and the Funda\c{c}\~ao de Apoio \`a Pesquisa do Distrito Federal (FAPDF).
H.B.\ acknowledges CNPq (Grant No.~308402/2022-7). K.Z.\ acknowledges CNPq (Grant No.~305665/2025-1).
\end{acknowledgments}

\appendix

\section{Pseudo-critical point of the spin-1/2 chain}
\label{app:pseudocritical}

\begin{figure}[h]
  \centering
  \includegraphics[width=\linewidth]{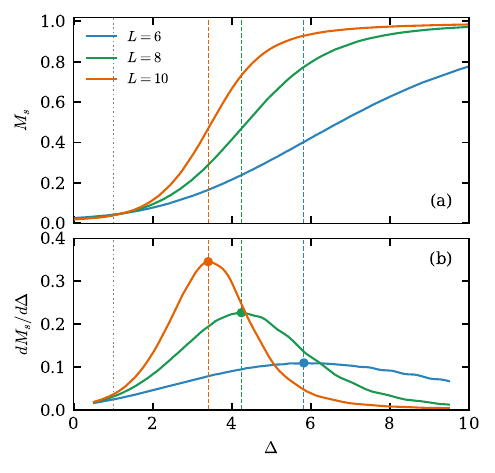} 
  \caption{(a) Staggered magnetization $M_s$ of the ground state of the
  spin-1/2 chain with $h = 0.1$ and (b) its derivative $dM_s/d\Delta$, as
  functions of $\Delta$, for $L = 6$, $8$, and $10$. The circles and dashed lines
  mark the maximum of $dM_s/d\Delta$, which defines the pseudo-critical point
  $\Delta^*(L)$, and the dotted lines mark the BKT transition at $\Delta = 1$.}
  \label{fig:pseudocritical}
\end{figure}

Figure~\ref{fig:pseudocritical}(a) shows the staggered magnetization of
the ground state for $L=6$, $8$, and $10$. Its derivative $dM_s/d\Delta$
plays the role of a susceptibility, and its maximum defines the
pseudo-critical point $\Delta^*(L)$~\cite{fisher1972scaling,barber1983finite}. It lies
at $\Delta^*\approx5.8$, $4.2$, and $3.4$ for $L=6$, $8$, and $10$,
respectively, Fig.~\ref{fig:pseudocritical}(b), sharpening and moving
toward $\Delta=1$ as $L$ increases, but remaining far from it, as expected
from the slow convergence of pseudo-critical points at a BKT transition.

\section{Dependence on the pinning field}
\label{app:pinning}

\begin{figure}[t]
    \centering
    \includegraphics[width=\columnwidth]{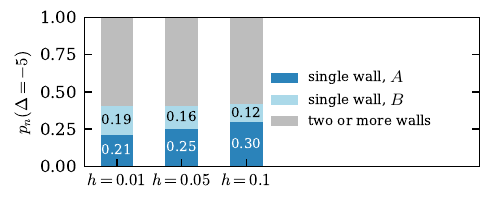}
    \caption{Populations of the evolved state at the end of the negative
    ramp of the spin-$1/2$ chain with $L=10$, for $\tau=35$ and different
    pinning fields $h$: single-wall state $A$, favored by the field,
    single-wall state $B$, and states with two or more domain walls.}
    \label{fig:pinning}
\end{figure}

Figure~\ref{fig:pinning} shows how the population of the evolved state at the end of the negative ramp is distributed among the domain-wall states, for $\tau=35$ and three values of the pinning field. The total population of the single-wall states, and hence that of the states with two or more walls, is nearly independent of $h$. The distribution between $A$ and $B$, in contrast, depends on $h$, and approaches the equal weights imposed by spin-flip symmetry as $h\to0$.

\section{Finite-size effects in the ramp dynamics}
\label{app:finite_size}

\begin{figure}[t]
\centering
\includegraphics[width=\columnwidth]{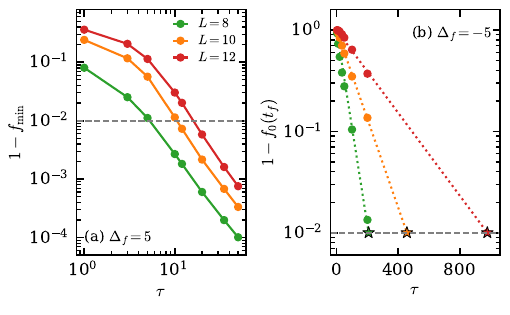}
\caption{Loss of fidelity as a function of $\tau$ for the spin-1/2 chain with $h = 0.1$ and $L = 8$, 10, and 12: (a) maximum loss of fidelity during the positive ramp, to $\Delta_f = 5$, and (b) loss of fidelity at the end of the negative ramp, to $\Delta_f = -5$. The dotted lines in (b) are fits $1 - f_0(t_f) = e^{-\tau/\tau_0}$, extended to $1 - f_0(t_f) = 10^{-2}$; the stars mark the extrapolated values of $\tau_{\rm ad}$. The dashed lines mark $1 - f_{\min} = 10^{-2}$ and $1 - f_0(t_f) = 10^{-2}$.}
\label{fig:finite_size}
\end{figure}

To assess how the results depend on the chain length, we repeated the ramps of the spin-1/2 chain for $L = 8$, 10, and 12, with $h = 0.1$. Figure~\ref{fig:finite_size} shows the loss of fidelity for the positive and negative ramps.

For the positive ramp, Fig.~\ref{fig:finite_size}(a), the loss of fidelity $1 - f$, evaluated at the minimum of $f$ during the ramp, grows with $L$ at fixed $\tau$. The adiabatic ramp time defined in Sec.~\ref{sec:characterization} grows only slowly, from $\tau_{\rm ad} \approx 5$ for $L = 8$ to $\tau_{\rm ad} \approx 10$ and $11$ for $L = 10$ and 12.

The negative ramp is much more sensitive to the chain length, Fig.~\ref{fig:finite_size}(b). The weight outside the single-wall state $A$ keeps the exponential form $1 - f_0(t_f) \simeq e^{-\tau/\tau_0}$ for all $L$, but $\tau_0 \approx 49$, 101, and 184 for $L = 8$, 10, and 12, roughly doubling every two sites, consistent with a minimum gap within the $S^z_{\rm tot} = 0$ sector near $\Delta = -1$ that decreases with $L$. Extrapolating the fits to the adiabaticity criterion gives $\tau_{\rm ad} \approx 210$, 460, and 870, so that the ratio between the adiabatic ramp times of the negative and positive ramps grows from about 40 for $L = 8$ to about 80 for $L = 12$. The qualitative picture of the main text is thus unchanged for the sizes considered, but the separation between the timescales of the first-order and BKT transitions increases with the system size.

\bibliography{refer}

\end{document}